**Title:**

# Miniaturized Circuitry for Capacitive Self-sensing and Closed-loop Control of Soft Electrostatic Transducers

**Authors:**

Khoi Ly,[1] Nicholas Kellaris,[1,2] Dade McMorris,[3] Brian K. Johnson,[1] Eric Acome,[1] Vani Sundaram,[1] Mantas Naris,[1] J. Sean Humbert,[1] Mark E. Rentschler,[1] Christoph Keplinger,[1,2] Nikolaus Correll[4]

**Affiliations:**

[1]Department of Mechanical Engineering, University of Colorado, Boulder, Colorado, USA.

[2]Materials Science and Engineering Program, University of Colorado, Boulder, Colorado, USA.

[3]Department of Electrical Engineering, University of Colorado, Boulder, Colorado, USA.

[4]Department of Computer Science, University of Colorado, Boulder, Colorado, USA.

**Abstract—** Soft robotics is a field of robotic system design characterized by materials and structures that exhibit large-scale deformation, high compliance, and rich multifunctionality. The incorporation of soft and deformable structures endows soft robotic systems with the compliance and resiliency that makes them well-adapted for unstructured and dynamic environments. While actuation mechanisms for soft robots vary widely, soft electrostatic transducers such as dielectric elastomer actuators (DEAs) and hydraulically amplified self-healing electrostatic (HASEL) actuators have demonstrated promise due to their muscle-like performance and capacitive self-sensing capabilities. Despite previous efforts to implement self-sensing in electrostatic transducers by overlaying sinusoidal low-voltage signals, these designs still require sensing high-voltage signals, requiring bulky components that prevent integration with miniature, untethered soft robots. We present a circuit design that eliminates the need for any high-voltage sensing components, thereby facilitating the design of simple, low cost circuits using off-the-shelf components. Using this circuit, we perform simultaneous sensing and actuation for a range of electrostatic transducers including circular DEAs and HASEL actuators and demonstrate accurate estimated displacements with errors under 4%. We further develop this circuit into a compact and portable system that couples HV actuation, sensing, and computation as a prototype towards untethered, multifunctional soft robotic systems. Finally, we demonstrate the capabilities of our self-sensing design through feedback-control of a robotic arm powered by Peano-HASEL actuators.

# MAIN TEXT

## Introduction

Traditional, rigid-bodied robots have high efficiency and precision in a well-defined environment but have limited capabilities in environments with uncertainty. On the other hand, soft robots exhibit large-scale deformation, high compliance, and the rich multifunctionality of natural organisms.[1-4] Such characteristics represent several advantages, including the ability to mitigate uncertainty with passive compliance,[5] perform highly dexterous tasks,[6] and exhibit resilience toward impacts, wear, and tear.[7] Soft robotics is a growing field of study, with numerous methods of actuation including thermal actuation,[8,9] pneumatic actuation,[10-14] hydraulic actuation,[15] cable-driven actuation,[8] and electrostatic actuation.[16-20] Of these, electrostatic transducers such as dielectric elastomer actuators (DEAs) and electrohydraulic HASEL actuators are particularly attractive due to their high-speed operation, high efficiency, controllability, and the ability to produce muscle-like forces and strains.[16-20]

In the past, there have been numerous efforts to sense and control electrostatic transducers. Several capacitance sensing methods have been developed, including periodic high voltage (HV) input,[21] transient effects of charging and discharging,[22] and changes in resistance in the transducer electrodes.[23] The first two sensing methods are unable to sense and drive the transducers independently, while the third method is susceptible to signal creep due to a slow decrease in electrode resistance over time. One of the most common capacitive self-sensing techniques for DEAs is based on the concept of superposition, or coupling, of a sinusoidal low voltage (LV) signal for sensing on a HV output signal for driving the transducer.[24-27] In these designs, the LV signal is coupled to the HV driving signal on the HV side of the electrostatic transducers, therefore the circuit designs require large, specialized, or expensive electronic components for combining the sinusoidal LV signals with the driving HV signals. This approach is limited to either using a HV power amplifier capable of high frequency signal generation or including an AC-AC transformer with sufficient HV isolation to protect the LV sensing components. To address the limitation of superimposing a sinusoidal waveform on a HV driving signal, Landgraf et al. introduced a self-sensing method in which the output of the sinusoidal voltage generator was applied to the LV side of the DEA.[28] While this method eliminated the need of a specialized waveform generator, the measurement method used in their design still occurred on the HV side of the circuit. As a result, the estimation of the DEA's capacitance still relied on an external, specialized measuring device that could probe the HV across the measuring resistor.

*Contributions*

This paper introduces a new method of capacitive self-sensing where both the generation and measurement of the LV sensing signal occur on the LV side of an electrostatic transducer. This implementation allows the transducer to inherently provide HV isolation for the sensing components, irrespective of the HV amplitude required for a given transducer, thus providing an application agnostic setup. This method further incorporates low impedance paths to ground through the sensing hardware, which allows for robust operation even in the event of electrical failure (e.g. dielectric breakdown) of the transducer. Therefore, this method eliminates the need for any high-voltage sensing components, and facilitates the design of a simple, low-cost circuit using off-the-shelf integrated components (Figure 1a-b). Using this method, we can simultaneously deform (either through HV or mechanical manipulation) and estimate the capacitance for a wide range of electrostatic transducers including a circular DEA and HASEL actuators, which can be readily mapped to an estimated displacement of the transducer. We also demonstrate the capabilities of the circuit with a feedback-controlled HASEL-powered arm (Figure 1c). This study is the proof of concept for a miniature system that couples actuation, sensing, and computation for a wide range of electrostatic transducers powered by a variety of HV supplies.

**Materials and Methods**

*Electrical model of a soft electrostatic transducer*

An electrostatic transducer can be electrically modeled as a variable capacitor with a series resistance and a parallel resistance (Figure 1d). As the transducer deforms under the influence of electrical or mechanical forces, its capacitance changes. The series resistance represents the resistance of the electrodes, which can be variable or fixed depending on the electrode design; in particular, the stretchable electrodes used in DEAs often exhibit resistances that are strongly dependent on deformation of the conductive percolating networks that occurs during actuation,[29] or degradation of the conductive networks with repeated cycling.[30] The variable parallel resistance represents the resistance of the dielectric material, which changes as the transducer deforms.

*Low voltage coupling self-sensing method*

The self-sensing method presented here measures the capacitance of an electrostatic transducer during deformation using 1) a LV sinusoidal signal coupled through the LV side of an electrostatic transducer and 2) measuring changes in sensing signals also on the LV side of the transducer. Figure 1d and Figure S1 show a schematic of the self-sensing circuit, transducer model, and a generalized HV power supply. Similar to previous works, our capacitive self-sensing method is based on measuring the voltage drop across the measuring resistor ($V_\mathrm{m}$) and its phase difference compared to the net sinusoidal sensing voltage signal ($V_\mathrm{net}$). Unlike the previous design,[28] we connected the measuring resistor ($R_m$) to the LV side of the transducer such that the voltage drop across the measuring resistor ($V_\mathrm{m}$) can be measured with components that are both off-the-shelf and integrated into the miniaturized self-sensing hardware, making the system self-contained.

In our circuit design, the HV driving signal is applied to one electrode and the high-frequency sinusoidal sensing voltage signal is applied to the other electrode of the transducer. The sensing LV signal is therefore coupled to the HV driving signal via the low-voltage side of the electrostatic transducer. The sinusoidal LV signal takes advantage of the internal path to ground through the HV supply via its HV ripple filtering capacitor ($C_f$) and its parallel resistor ($R_{pp}$). From the schematics in Figure 1d and Figure S1, the transducer ($C_a$) and the HV ripple filtering ($C_f$) capacitors are connected in series and the combined component is placed in parallel with a transient voltage suppression capacitor ($C_t$). Since virtually all power supplies have a HV ripple filtering capacitor ($C_f$) the proposed design can be applied to a wide range of HV supplies; in the absence of a filtering capacitor, a separate path to ground can be added, such as a resistor or capacitor external to the power supply. In the event of the transducer's electrical failure, the inclusion of the transient voltage suppression capacitor ($C_t$), together with a pair of Zener diodes (Figure 1d and Figure S1), helps suppress change in the junction voltage ($V_j$) with HV transient signals, and filters the HV ripple from the HV supply, thus protecting the sensing circuit and providing a reliable way to measure voltage across the measuring resistor. However, too large of a transient suppression capacitor decreases the self-sensing sensitivity, as it becomes a low impedance path to ground. Ideally, the transient suppression capacitor and the Zener diodes are not critical for the operation of the self-sensing circuit, as long as the transducer remains fully functional.

*Calculating transducer capacitance from self-sensing*

From Figure S1, we can calculate the capacitance of the electrostatic transducer; by choosing the sinusoidal LV sensing signal ($V_{net}$), measuring the voltage across the measuring resistor ($V_m$), and calculating their phase difference using a Discrete Fourier Transform algorithm, the transducer's capacitance can be calculated using AC signal analysis. The net current output from the sine wave generator, $I_{net}$, is the same as the current drawn by the measuring resistor, $I_m$, and equals the sum of currents through the transient suppression capacitor ($I_t$) and the transducer ($I_a$):

$$I_{net} = I_m = I_t + I_a. \tag{1}$$

It is assumed that the transducer has infinitely large parallel resistance ($R_p$); therefore, the current through the transducer ($I_a$) does not include the current through the parallel resistance. Similarly, the net voltage, $V_{net}$, equals the sum of the voltage drop across the measuring resistor ($V_m$) and the transient suppression capacitor, which is also the junction voltage ($V_j$):

$$\vec{V}_{net} = \vec{V}_m + \vec{V}_j. \tag{2}$$

Using Ohm's law, we obtain the relationship between the voltage across the measuring resistor and the impedance of the transducer:

$$\frac{\|V_\mathrm{m}\|}{R_\mathrm{m}} = \frac{\|V_\mathrm{j}\|}{Z_\mathrm{t}} + \frac{\|V_\mathrm{a}\|}{Z_\mathrm{a}}. \tag{3}$$

From these equations, we can calculate the capacitance of the electrostatic transducer ($C_\mathrm{a}$). Appendix A1 shows the detailed derivation of the capacitance. The final equation is as follows:

$$C_\mathrm{a} = \frac{1}{2\pi Z_{ca} f}, \tag{4}$$

Where

$$Z_\mathrm{ca} = \frac{\frac{V_\mathrm{a}}{I_\mathrm{a}} \sin(\beta) \left(\frac{1}{\tan(\beta)^2} + 1\right)}{2} + \frac{\sqrt{\frac{V_\mathrm{a}^2}{I_\mathrm{a}^2} \sin(\beta)^2 \left(\frac{1}{\tan(\beta)^2} + 1\right)^2 - 4R_\mathrm{s}^2}}{2}, \tag{5}$$

$Z_\mathrm{ca}$: Impedance of the variable capacitance of the electrostatic transducer ($\Omega$)
$C_\mathrm{a}$: Variable capacitance of the electrostatic transducer (F)
$f$: Frequency of sine wave generator (Hz)
$V_\mathrm{a}$: Voltage drop across the electrostatic transducer (V)
$I_\mathrm{a}$: Current across the electrostatic transducer (A)
$R_\mathrm{s}$: Electrode resistance of electrostatic transducer ($\Omega$)
$\beta$: Calculated phase lead between $I_\mathrm{a}$ and $V_\mathrm{a}$ (Rads)

*Complete system design for self-sensing*

To realize the self-sensing system, a direct digital synthesizer (AD9833, Linear Technology) is chosen to generate a 540-mV sinusoidal wave at 1 kHz to the self-sensing circuit network (Figure 1a and 1b). An inverting op amp (LT1037, Linear Technology) with a gain of 1:33.3 amplifies the signal to 24 V peak-to-peak. An instrumentation amplifier (INA129, Texas Instruments) is used to measure the voltage across the measuring resistor ($R_\mathrm{m}$). Its voltage output is attenuated 10:1, passed through a 2nd order low-pass filter with a 10 kHz corner frequency to filter the HV ripple from the HV amplifier, and level-shifted by +1.25 V. The presented hardware and the choice of the software filter allow the self-sensing circuit to reliably measure the transducers' capacitance at 96 Hz. Since the integrated analog-to-digital converter (ADC) of the microcontroller (MK66FN2M0VLQ18, NXP) can only accept 0 – 2.5 V, the signal conditioning circuit is used to scale the voltage across the measuring resistor into a measurable range. A 220-pF transient suppression capacitor is chosen to help the self-sensing circuit function well with a Trek 50/12 HV amplifier (Trek, Inc.) that has a maximum slew rate of 350 V/µs. Our system can reliably sense the capacitance even with a step HV input at 10 kV. We also implement a bidirectional Zener diode setup which is composed of two Zener diodes (DDZ9713S-7, Diodes Incorporated). In the event of high voltage discharge in the transducer (e.g. through dielectric breakdown), the Zener diodes protect several components connected to the junction from a sudden voltage increase. While the control and power PCB costs approximately $90, the self-sensing PCB only costs $50 using off-the-shelf components. This makes the presented self-sensing circuit a simple, versatile, cost effective, and compact system for sensing capacitance in electrostatic transducers. The detailed schematics of the processing and self-sensing circuits are shown Figure S2, S3, and S4.

**Experiments and Results**

The self-sensing circuit performs well for a variety of soft electrostatic transducers and sensors. In this section, we validate our self-sensing method using a dielectric elastomer actuator (DEA) and a capacitive stretchable sensor; we compare the self-sensing results to theoretical capacitance changes calculated from geometric models of each transducer.

*System validation using a dielectric elastomer actuator (DEA)*

The basic structure of a DEA includes an elastomeric dielectric layer that is sandwiched between stretchable electrodes. When a HV signal is applied, electrostatic forces cause the dielectric to decrease in thickness and expand in area. Circular DEAs, as shown in Figure 2a, are a common shape because the axisymmetric expansion that occurs during actuation is uniform and convenient to characterize. As shown in Appendix A2, the relative change in capacitance, $\Delta C/C_0$, is related to the stretch of the DEA, $\lambda_r$, in the radial direction as follows

$$\frac{\Delta C}{C_0} = \frac{C - C_0}{C_0} = \lambda_r^4 - 1. \tag{6}$$

The performance of our self-sensing circuit with a DEA is verified by measuring the self-sensing capacitance while simultaneously recording radial stretch using a Canon EOS 6D DSLR camera recording at 60 frames per second. The resulting video files were processed using a custom MATLAB program.[19]

The DEA was made from an acrylic elastomer (VHB 4910, 3M) that was pre-stretched radially 2.2 times ($\lambda_r = 2.2$) onto an acrylic ring. Then carbon grease electrodes were painted onto either side of the dielectric (846, MG Chemicals). The electrode diameter was approximately 20 mm. The actuation signal for the DEA was a sinusoidal waveform with constant frequency (0.5 Hz) and increasing amplitude (0 – 4.4 kV).

Figure 2b shows the calculated relative change in capacitance plotted along with capacitance measured using our self-sensing circuit. The capacitance measured with our sensor overlaps with the relative change in capacitance based on optical measurements, showing good agreement between the two methods with a root mean square error (RMSE) of 0.0299. Typical response of the capacitive self-sensing for the DEA is shown in Video S1.

*System validation using a capacitive stretch sensor*

A stretchable rectangular parallel-plate capacitive sensor is used to demonstrate the self-sensing capabilities of our circuit in the absence of a HV power supply (Figure 2c). As shown in Appendix A3, the relative change in capacitance, $\Delta C/C_0$, is linearly related to stretch in the y-direction, $\lambda_y$, for the capacitive sensor as follows

$$\frac{\Delta C}{C_0} = \lambda_y - 1 \tag{7}$$

To verify that our self-sensing circuit works with a stretchable sensor, we fabricated a soft 1 x 4 cm rectangular parallel-plate capacitor using 500 µm EcoFlex 00-30 for the dielectric and silver knitted fabric (Medtex P130, V Technical Textiles, Inc.) for the electrodes. The sensor, pre-stretched to 6.5 cm, was fixed at one end and the other end was mounted to a linear bearing so that the stretch was constrained only to the y-direction. A force, $F$, was applied to manually stretch the sensor in a random pattern. We measured capacitance using the self-sensing circuit and simultaneously monitored change in length of the sensor using a laser displacement sensor (LK-H157, Keyence Corp.).

Figure 2d shows relative change in capacitance plotted along with capacitance measured using our self-sensing circuit. The capacitance measured with our sensing circuit overlaps with the relative change in capacitance based on laser displacement measurements, showing good agreement between the two methods with an RMSE of 0.0048.

*Characterization of self-sensing with HASEL actuators*

Unlike DEAs and parallel-plate capacitive stretch sensors, HASEL actuators have more complex geometries,[19,20,31] heterogeneous multiphase structures, and they can exhibit certain types of instabilities in actuation,[32] which complicates the mathematical calculation of displacements with respect to changes of capacitance. Therefore, we propose a calibration procedure to map the displacement of the HASEL actuators and the change of their capacitances. This procedure would need to be performed for each unique design (e.g. materials and geometry) of HASEL actuator due to their distinct capacitance-displacement relationships. However, once the calibration curve is determined for a specific design of HASEL actuator, it can be applied without modification to all actuators of the same design.

Self-sensing calibration is dependent on the HV power supply and the transducer being used. Here we show the self-sensing calibration using a Trek 50/12 HV amplifier with two types of HASEL actuators: a linearly-contracting Peano-HASEL actuator (introduced by Kellaris et al.[20]) with eight pouches connected in series, and a linearly-expanding folded-HASEL actuator (introduced by

Mitchell et al.[31]) made from eleven pouches folded into a stack. Fabrication details for both types of HASEL actuators are presented in Appendix A4. The calibration procedure contains two parts: offline characterization (Figure 3), in which the self-sensing data are collected then later analyzed, and real-time validation (Figure 4), in which the displacements of the HASEL actuators are estimated based on real-time analysis of the self-sensing capacitance data.

In the offline characterization, the calculated capacitances of the Peano-HASEL actuator with a 200 g load and folded-HASEL actuator with a 150 g load are synchronized with both the laser (true) displacement data (LK-H157, Keyence Corp.) and the HV applied. The estimated displacements of the HASEL actuators are calculated using a second order polynomial fitting function (MATLAB R2019b, MathWorks). The function is obtained from the total self-sensing capacitance data and the true displacement data that are collected from two set of experiments: one with sinusoidal HV frequencies from 0.1 Hz to 5 Hz, with 2.5 kV amplitude and 3.5 kV offset and the other with sinusoidal HV frequency at 0.5 Hz, with 0.5 kV to 2.5 kV amplitude and 3.5 kV offset. For each HV sine wave, the data are collected 10 times, with the error bars as one standard deviation from the mean of 10 repeated experiments.

As can be seen from Figure 3, the self-sensing of the Peano-HASEL (Figure 3a) and folded-HASEL actuators (Figure 3b) using a Trek 50/12 shows good agreement between the true displacements and estimated displacements in the entire HV amplitude range of interest (Figure 3c-d). The scatter plots of estimated versus true displacements based on the calibration data for Peano-HASEL and folded-HASEL actuators show excellent goodness of fit (R-squared) values of 0.992 and 0.987, respectively (Figure 3e-f). The RMSE between estimated and true displacement, normalized by the range of displacement for each actuator to give NRMSE, is shown for Peano-HASEL and folded-HASEL actuators as a function of HV amplitudes (Figure 3g) and HV signal frequencies (Figure 3h). For varying HV amplitudes, the average NRMSE is only 0.04 for both Peano-HASEL and folded-HASEL actuators (Figure 3g). Similarly, the self-sensing performs well for actuation frequencies up to 1 Hz, with NRMSE values less than 0.10 for both actuators (Figure 3h). Beyond frequencies of 1 Hz, the self-sensing performance decreases due to a phase lag between the self-sensing signal and the change in actuator displacement (Figure 3i). While the reason behind the phase lag is still unknown, Figure S5 suggests that the issue lies in the HASEL actuators, rather than the self-sensing circuit (which samples at 96Hz); the sinusoidal displacement for the parallel-plate stretchable capacitor, with increasing frequencies from 0.5 Hz to 3 Hz, does not cause any phase lag between the change in self-sensing capacitances and relative change in theoretical capacitance. It is possible that the non-linear electrode zipping of HASEL actuators causes a phase discrepancy in the resultant electromechanical dynamics.

*Validation of self-sensing with HASEL actuators*

With the second order polynomial calibration function programmed in the self-sensing microcontroller (Figure 3), the validation procedure demonstrates the agreement between the true and estimated displacements of the Peano-HASEL and folded-HASEL actuators in real-time, as shown in Figure 4. For the Peano-HASEL actuator, the self-sensing system shows good agreement between the estimated and true displacement data with RMSEs of 0.05 mm for a HV driving signal with both increasing amplitude and frequency (Figure 4a), 0.07 mm for a HV signal with constant amplitude and increasing frequency (Figure 4c), and 0.04 mm for a HV driving signal with decreasing amplitude and increasing frequency (Figure 4e). The percent errors are 1.81%, 1.95%, and 1.50%, respectively. It is consistent with the characterization results that there are expected mismatches in displacements at frequencies higher than 1 Hz (Figure 4c), but the mismatches disappear when the displacement is less than 2.5 mm (Figure 4a), even if the actuation frequency is above 1 Hz (Figure 4e). For the folded-HASEL actuator, the RMSEs between the estimated and true displacement data are 0.03 mm for HV signal with both increasing amplitude and frequency (Figure 4b), 0.03 mm for HV signal with constant amplitude and increasing frequency (Figure 4d), and 0.02 mm for HV signal with decreasing amplitude and increasing frequency (Figure 4f). The percent errors are 3.33%, 3.36%, and 2.58%, respectively.

*Sensitivity demonstration of self-sensing*

To qualitatively demonstrate the self-sensing sensitivity, 5 cycles of a 0.05 Hz trapezoidal HV waveform from 0 kV to 5.5 kV were applied to a Peano-HASEL actuator. The ability of the self-sensing circuit to detect the change in capacitance of the actuator under various types of disturbances can be seen in Figure 5 and Video S2. The driving HV has an approximately 40 kHz ripple, and noise with frequencies from 0 Hz to 2 kHz. We believe that the HV noise is dependent on the capacitive load. Therefore, self-sensing with driving HV turned on has a sensitivity that is dependent on the HV values and the zipping state of the actuator. When there is no HV, the calculated capacitance has a peak-peak noise of 1 pF, and the sensing system can detect a 20 g load. With HV signal turned on and 0 g load, the calculated capacitance values fluctuate with a 100-pF peak-peak noise, but it decreases to 50 pF when there is a 100 g load applied.

Physical interactions such as pulling on the Peano-HASEL actuator also result in corresponding changes in capacitance. The self-sensing circuit can detect a single pull from $t = 65$ s to $t = 75$ s and rapid pulls from $t = 80$ s to $t = 90$ s (Video S2). The self-sensing method is sufficiently robust to detect charge retention in the Peano-HASEL actuator, which occurs after applying high electric fields for a period of time and causes full actuator relaxation (lowest capacitance) at low voltages of ~ 1 kV, and a slight reactivation (e.g. contraction and slightly higher capacitance) at 0 kV. While this effect is not well understood, it may be related to space charge creation in the thin-film dielectrics, caused by ionization of low-molecular-weight impurities under application of high electric fields.[33,34] This dielectric loss, while affecting the actuators' response to HV application, does not influence the measurements from self-sensing. Rather, the self-sensing circuit can even qualitatively detect the degree of space charge accumulation in the thin-film dielectrics based on the actuators' elevated capacitances under no HV, as shown in Video S2.

*Self-sensing demonstration using a miniature HV DC-DC converter*

In this section we present a portable self-sensing system using an off-the-shelf miniature HV DC-DC converter (EMCO C80, XP Power) with a Peano-HASEL actuator. The ability to use of an unmodified, 8-cm long miniature converter instead of the 1.5 m tall Trek 50/12 (used in Figure 2a, and Figures 3-5) with the self-sensing circuit opens many possibilities in portable, untethered HV soft robotic designs.

Figures 6a-b show the physical representation of the portable driving and self-sensing system. A driving sinusoidal HV waveform at 0.1 Hz was applied to the actuator with increasing amplitude (0 kV to 1.5 kV), constant amplitude (1.5 kV) and decreasing amplitude (1.5 V to 0 kV) with 2.3 kV offset. The choice of HV amplitudes and frequencies provides visualization of how self-sensing circuit performances within the HV range of interest when using the EMCO C80 converter. The lower bound and upper bound of the HV signal acount for the actuators' activation voltage around 1 kV and the distortion of the self-sensing signal at 3.3 kV. The offline estimated displacements – based on a $2^{nd}$ order polynomial fitting function for the three cases of HV waveforms – are captured and shown in Figure 6c-e. As shown in Figure 6e, the self-sensing capacitance is distorted for HV signals above 3.3 kV when using the EMCO C80 instead of the Trek 50/12. We believe that the increase of HV beyond 3.3 kV results in larger interfering noise at the same frequency but out of phase with the sensing signal. The cancelation of the HV output noise and the self-sensing sine wave results in the distorted capacitance readouts. The problem could be mitigated significantly by increasing the amplitude of the self-sensing waveform to increase the signal-to-noise ratio. However, the miniature self-sensing and true displacements agree well with RMSE of 0.04 mm (2.92% error), 0.07 mm (3.16% error), and 0.05 mm (3.15% error) for the HV waveforms with increasing amplitude, constant amplitude, and decreasing amplitude, respectively.

*Closed-loop control with Self-sensing circuit*

In this section, we demonstrate the ability to track a variety of displacement reference waveforms using a proportional-integral-derivative (PID) controller with the calibrated self-sensing displacements as feedback. As the EMCO C80 has significant asymmetric 300-ms rise and 3-s fall times, irrespective of the loads, we opted to demonstrate the closed-loop control with the Trek 50/12. However, it is noted that better miniature HV amplifiers, either off-the-shelf or custom, could be used with the presented self-sensing circuit for portable, untethered soft-robotic applications.

The plant to be controlled is a robotic arm driven by two Peano-HASEL actuators lifting 20-g load (Figure 1c). The self-sensing capacitance-to-displacement calibration results for this specific plant are shown in Figures 7a-c. The closed-loop controller has a frequency of 90 Hz. The real-time estimated self-sensing displacement is filtered by a first-order low-pass digital filter with 5 Hz corner frequency. The PID controller was tuned experimentally; the chosen proportional, integral, and derivative gains are: $K_P = 0.045$, $K_I = 0.9$, and $K_D = 0.001$. The PID algorithm computed the desired high voltage value to reduce the errors between the target displacements and the self-sensing displacements. This high voltage value was mathematically scaled down by a factor of 1/5000 so that it could be generated by a low voltage digital to analog converter from the microcontroller. The scaled analog output voltage was then amplified by the Trek 50/12 with an amplification of 1:5000, thus generating a proper HV value to adjust the plant's displacements.

Figure 7d shows a typical tracking performance of the closed-loop system with respect to three displacement targets: steps, ramp, and sine waveforms. The tracking performance is satisfactory with 1.00% and 1.70% steady state errors between the true displacement and the target displacement at 16 mm, 24 mm steps, respectively. The controller has excellent rise times of 267 ms and 100 ms for a 4-to-16-mm step and a 16-to-24-mm step, respectively. The corresponding fall times are slightly worse with 340 ms and 120 ms for a 16-toF-4-mm step and 24-to-16-mm step, respectively. The asymmetric step responses of the HASEL arm can be explained by the asymmetric activation and relaxation characteristics of the Peano-HASEL actuators. The activation process is an active force application process due to the HV being applied directly to the dielectric liquid via the electrode zipping. In contrast, relaxation is a passive restoration process where the electrodes unzip as the result of the viscoelastic return flow of the FR3 dielectric liquid and the lack of electrostatic force. The ramp tracking result is satisfactory with small errors of 3.97% when HV signal is increasing and 3.56% when HV signal is decreasing. The sine tracking at 0.2 Hz is also good with 7.92º phase lag between the true displacement and the target displacement. A real-time demonstration of the reference displacement tracking is shown in Video S3. Overall, the self-sensing circuit has proven its capability for controlling HV electrostatic transducers.

**Conclusion**

In this paper, we introduce a new capacitive self-sensing method that enables simultaneous actuation and estimation of the deformation of a wide range of electrostatic transducers. Using the transducer to provide HV isolation, applying a sensing signal on the low-voltage side of the transducer, and connecting the measuring resistors to the LV side of the transducer enable a simple, robust, and miniaturized circuit design which uses inexpensive, off-the-shelf, and integrated components and therefore potentially accommodates a wide range of HV DC-DC converters. The performance of the self-sensing circuit is quantitatively validated using a circular DEA and a stretch sensor. The versatility of the circuit is also demonstrated using two different HASEL actuators (Peano-HASELs and folded-HASELs) and two different HV amplifiers (Trek 50/12 and EMCO C80). As an application of the self-sensing circuit, we demonstrate the feedback-control ability using a PID controller for a robotic arm powered by Peano-HASELs. While the high slew rate of the wall-powered Trek 50/12 HV amplifier allows much faster closed-loop control, the EMCO C80 DC-DC converter and self-sensing circuit demonstrates a fully miniaturized, self-contained system with integrated actuation, sensing, and computation capabilities.

Moving forward, there are many opportunities to further explore the self-sensing performance of this circuit. Instead of using an integrated ADC in the microcontroller with a maximum sampling rate of 40 kHz, the sampling rate of the self-sensing signal can be greatly improved using a dedicated high-speed ADC with direct memory access (DMA) for much faster sampling rate, possibly up to 1 MHz. This will allow the self-sensing sine wave frequency to increase to 10 kHz, potentially increasing the capacitance update rate to 1 kHz. In addition, self-sensing can be a great tool to monitor the electrostatic transducer's electrical integrity. A soft electrostatic transducer has a significantly smaller parallel resistance after experiencing dielectric breakdown, and therefore generates a noisy signal on the self-sensing output. The self-sensing circuit can withstand the dielectric breakdown events and detect transducer failure by capturing the spurious noises from the measured capacitance. The effects of dielectric losses have not been investigated in this work; however, the self-sensing method introduced here demonstrated the ability to detect the influence of space charge accumulation in HASEL actuators, accurately reflecting changes in strain due to this effect. Future work could probe the effects of other nonlinear dielectric processes such as the loss tangent (tan δ) on the fidelity of the self-sensing signal. Overall, the versatility and robustness of the self-sensing circuit introduced in this work promises many possible applications and implementations in soft robotic systems.

**Acknowledgements**

This research has been supported by the National Science Foundation CPS program (Grant No. 1739452), the Air Force Office of Scientific Research (Grant No. 83875-11094), and a GAANN (Graduate Assistantships in Areas of National Need) Fellowship in Soft Materials. The authors also acknowledge funding from the Army Research Office (Grant No. W911NF-18-1-0203), which was used to purchase laboratory equipment to characterize and fabricate transducers. We also acknowledge Dr. VP Nguyen and Dr. Eric Bogatin for their discussions regarding the PCB design of the self-sensing system.

**Author Disclosure Statement**

E.A., N.K., and C.K. are listed as inventors on a U.S. provisional patent application (62/638,170) and PCT application (PCT/US2018/023797) which cover fundamentals and basic designs of HASEL actuators. E.A., N.K., and C.K. are co-founders of Artimus Robotics, a start-up company commercializing HASEL actuators. M.R. is a co-founder of Aspero Medical, Inc., a start-up company that is focused on commercializing balloon overtube products for use in enteroscopy. N.

Address correspondence to:
Khoi Ly

Department of Mechanical Engineering
University of Colorado Boulder
Boulder, CO 80309
USA

E-mail: Khoi.Ly@colorado.edu

**Appendix**

**Appendix A1: Calculation of Self-sensing Capacitance**

Knowing the net voltage ($V_{net}$), the voltage across measuring resistor ($V_m$), and their phase difference ($\theta$), we can calculate the following:

$$V_j = \sqrt{V_{net}^2 + V_m^2 - 2V_{net}V_m\cos(\theta)}, \tag{1}$$

where $V_j$ is the voltage at the circuit junction connecting the measuring resistor, the transducer, and the HV transient suppression capacitor. The angle between the voltage drop across the measuring resistor and the junction voltage is then:

$$\varphi = \pi - \Psi, \tag{2}$$

$$\Psi = \sin^{-1}\left(\frac{V_{net}}{V_j}\cos(\theta)\right), \tag{3}$$

where $\Psi$ is the angle between the current through the measuring resistor and the junction voltage. The angle between the current through the power supply filtering capacitor and the current through the measuring resistor is:

$$\zeta = \frac{\pi}{2} - \Psi. \tag{4}$$

The currents through the measuring resistor ($I_m$), the HV transient suppression capacitor ($I_t$), and the electrostatic transducer ($I_a$) are:

$$I_m = \frac{V_m}{R_m}, \tag{5}$$

$$I_t = V_j(2\pi C_t f), \tag{6}$$

$$I_a = \sqrt{I_m^2 + I_t^2 - 2I_m I_t \cos(\theta)}. \tag{7}$$

$f$: Frequency of sine wave generator (Hz)

$C_t$: HV transient suppression capacitor (F)

From (4), (6), and (7), we can obtain the angle between $I_m$ and $I_a$:

$$\chi = \sin^{-1}\left(\frac{I_t}{I_a}\sin(\zeta)\right). \tag{8}$$

From (4) and (8), the angle α between $I_a$ and $V_j$ is then:

$$\alpha = \frac{\pi}{2} - \zeta - \chi. \tag{9}$$

Since current through the actuator is the same as current through the HV DC-DC converter, the amplitude of the sinusoidal voltage at the HV DC-DC converter's output line ($V_o$) is:

$$V_o = I_a\left(\frac{R_{pp}\sqrt{R_{ps}^2 + Z_c^2}}{R_{pp} + \sqrt{R_{ps}^2 + Z_c^2}}\right). \tag{10}$$

$R_{ps}$: Estimated series resistance of the HV DC-DC converter (Ω)

$R_{pp}$: Estimated parallel resistance of the HV DC-DC converter (Ω)

$Z_c$: Impedance of the HV filtering capacitor $C_f$ (Ω)

The angle between the sinusoidal voltage at the HV DC-DC converter's output line ($V_o$) and voltage across the HV filtering capacitor ($V_c$) is

$$\gamma = \tan^{-1}\left(\frac{R_{ps}}{Z_c}\right). \tag{11}$$

The angle between the current through the series resistance of the HV DC-DC converter ($I_{ps}$) and the voltage drop across the HV filtering capacitor ($V_c$) is

$$\omega = \frac{\pi}{2} - \gamma. \tag{12}$$

Since $V_o = V_{pp}$, The current through $R_{pp}$ equals to

$$I_{pp} = \frac{V_o}{R_{pp}}. \tag{13}$$

From (7), (12), and (13) the angle between the voltage at the HV DC-DC converter's output line ($V_o$) and the net voltage, $V_{net}$, is:

$$\rho = \omega - \sin^{-1}\left(\frac{I_{pp}}{I_a}\sin(\pi - \omega)\right). \tag{14}$$

The angle between the sinusoidal voltage at the HV DC-DC converter's output line ($V_o$) and the junction voltage ($V_j$) is

$$\eta = \rho - \alpha. \tag{15}$$

The voltage drop across the electrostatic transducer ($V_a$) is

$$V_a = \sqrt{V_j^2 + V_o^2 - 2V_jV_o\cos(\eta)}. \tag{16}$$

The angle between the voltage drop across the transducer and the current through it is

$$\beta = \alpha - \sin^{-1}\left(\frac{V_o}{V_a}\sin(\eta)\right). \tag{17}$$

The impedance of the capacitance of the electrostatic transducer is

$$Z_{ca} = \frac{\frac{V_a}{I_a}\sin(\beta)\left(\frac{1}{\tan(\beta)^2} + 1\right)}{2} + \frac{\sqrt{\frac{V_a^2}{I_a^2}\sin(\beta)^2\left(\frac{1}{\tan(\beta)^2} + 1\right)^2 - 4R_s^2}}{2}. \tag{18}$$

where $R_s$ is the electrode resistance of the electrostatic transducer. The self-sensing capacitance is therefore

$$C_a = \frac{1}{2\pi Z_{ca} f}. \tag{19}$$

**Appendix A2: Capacitance model for the tested DEA**

The change in shape that occurs during actuation results in a large charge in capacitance. Initial capacitance for a circular DEA is equal to,

$$C_0 = \varepsilon_0 \varepsilon_r \pi \frac{r_0^2}{t_0}, \tag{20}$$

where $\varepsilon_0$ is permittivity of vacuum (8.85 x $10^{-12}$ F/m), $\varepsilon_r$ is relative permittivity of the dielectric, $r_0$ is initial radius of the electrodes, and $t_0$ is initial thickness of the dielectric layer separating the electrodes. For axisymmetric radial expansion of an incompressible material,

$$\lambda_r^2 \lambda_z = 1, \tag{21}$$

where $\lambda_r$ is stretch in the radial direction ($r/r_0$) and $\lambda_z$ is stretched in the z-direction ($t/t_0$). When the DEA is actuated, the capacitance increases and is equal to

$$C = \varepsilon_0 \varepsilon_r \pi \frac{r_0^2}{t_0} \frac{\lambda_r^2}{\lambda_z}. \tag{22}$$

Using (21), the capacitance simplifies to

$$C = \varepsilon_0 \varepsilon_r \pi \frac{r_0^2}{t_0} \lambda_r^4. \tag{23}$$

Assuming an ideal dielectric, where permittivity does not change with stretch or applied electric field, the relative change in capacitance is equal to

$$\frac{\Delta C}{C_0} = \frac{C - C_0}{C_0} = \lambda_r^4 - 1. \tag{24}$$

**Appendix A3: Capacitance model for parallel-plate stretch sensor**

For this geometry, the initial capacitance is equal to,

$$C_0 = \varepsilon_0 \varepsilon_r \frac{l_0 w_0}{t_0}, \tag{25}$$

where $l_0$ is initial length of the electrodes, $w_0$ is initial width of the electrodes, and $t_0$ is initial thickness of the dielectric layer separating the electrodes. For uniaxial stretch in the y-direction of an incompressible material, we assume that stretch is equal in the x- and z-direction

$$\lambda_x = \lambda_z, \tag{26}$$

and capacitance when stretched is

$$C = \varepsilon_0 \varepsilon_r \frac{l_0 w_0}{t_0} \frac{\lambda_y \lambda_x}{\lambda_z}. \tag{27}$$

Using (26), capacitance when stretched simplifies to

$$C = \varepsilon_0 \varepsilon_r \frac{l_0 w_0}{t_0} \lambda_y, \tag{28}$$

which is proportional only to stretch in the y-direction for an ideal dielectric. The resulting change in capacitance is then equal to

$$\frac{\Delta C}{C_0} = \lambda_y - 1. \tag{29}$$

**Appendix A4: Fabrication of HASEL actuators**

Linearly-contracting Peano-HASEL actuators and expanding folded-HASEL actuators were fabricated using the method described in detail by Mitchell et al.[31] The fabrication procedure is outlined below.

*Peano-HASEL fabrication*

Two sheets of 12-um-thick polyester film (Mylar 850H, DuPont Teijin) were sealed into pouches using a modified commercially-available CNC (Carbide 3D Shapeoko XL) with a heated 3D printing extruder tip. A heating tip temperature of 195 °C with a tip sealing speed of 400 mm/min was used to seal the films. The films had a heat-sealing layer on one side; these heat-sealing layers faced inward during sealing to facilitate thermal bonding. A 25-um-thick layer of Kapton film was placed over the films during sealing to distribute heat and avoid damage from the heated tip. The pouch shape was drawn using CAD software (Solidworks 2019), exported to a dxf file, then converted to gcode using open source software (dxf2gcode). The Peano-HASEL actuators consisted of 8 pouches, with each pouch being 51mm wide by 16 mm long (Figure S6a). The bottom half of each pouch was notched to reduce side constraints during actuation. During sealing, a small fill port was left open on each pouch for later filling with liquid dielectric. After sealing, electrodes were deposited on the outer layer of the films using screen-printing. Electrodes were made from a carbon-based stretchable ink (CI-2051, Engineered Materials Systems, Inc.). Electrodes covered the full pouch width and half of the pouch length, as shown in Figure S6b. A small 1 mm margin was left between the edge of the electrodes and the heat seals to prevent dielectric breakdown through the seals during operation. Finally, the pouches were filled with 1.85 mL of a liquid dielectric (Envirotemp FR3, Cargill) using a syringe and needle inserted into the fill port of the pouch. Large bubbles were manually removed from each pouch, then a soldering iron (Weller WES51) heated to 195 °C was used to seal the fill port. A 25-um-thick layer of Kapton film was placed between the film and soldering iron tip to distribute heat and avoid damage to the film during sealing. After fabrication, the linearly-contracting Peano-HASEL actuators were mounted on acrylic frames to aid in uniform force transmission during operation.

*Folded HASEL fabrication*

Folded-HASELs were made using the same method as Peano-HASEL actuators. Fabrication began by heat-sealing together two layers of 20-um-thick polyester film (L0WS, Multi-Plastics) with an inner heat-sealing layer and corona-treated outer layer for ink adhesion. These films were sealed using the same CNC and sealing conditions (195 ºC, 400 mm/min) as the Peano-HASEL actuators. The folded HASEL actuators consisted of 11 pouches, each 30 mm by 30 mm, as shown in Figure S7a (only six pouches are shown in the figure for simplicity). Small notches were added at the sides of the pouch to reduce constraints and facilitate actuation. For folded-HASELs, all pouches were initially connected for simultaneous filling with liquid dielectric, and a small fill port was placed at the top and bottom of the pouch to allow filling from either end (Figure S7a). Electrodes (CI-2051, Engineered Materials Systems, Inc.) were deposited on either side of the pouch in the pattern shown in Figure S7b using a screen-printing method. Pouches were filled with a liquid dielectric (Envirotemp FR3, Cargill) – 0.76 mL per pouch – using a needle and syringe. After filling, large bubbles were manually removed, and a soldering iron was used to seal the fill ports, as with Peano-HASEL actuators. After, the connectors between each pouch were sealed off to isolate the liquid dielectric in each pouch. Finally, the pouches were folded together, alternating back and forth, along the fold points shown in Figure S7b. Small strips of two-sided adhesive (3M 924 transfer tape) were placed between each pouch to secure the stack together.

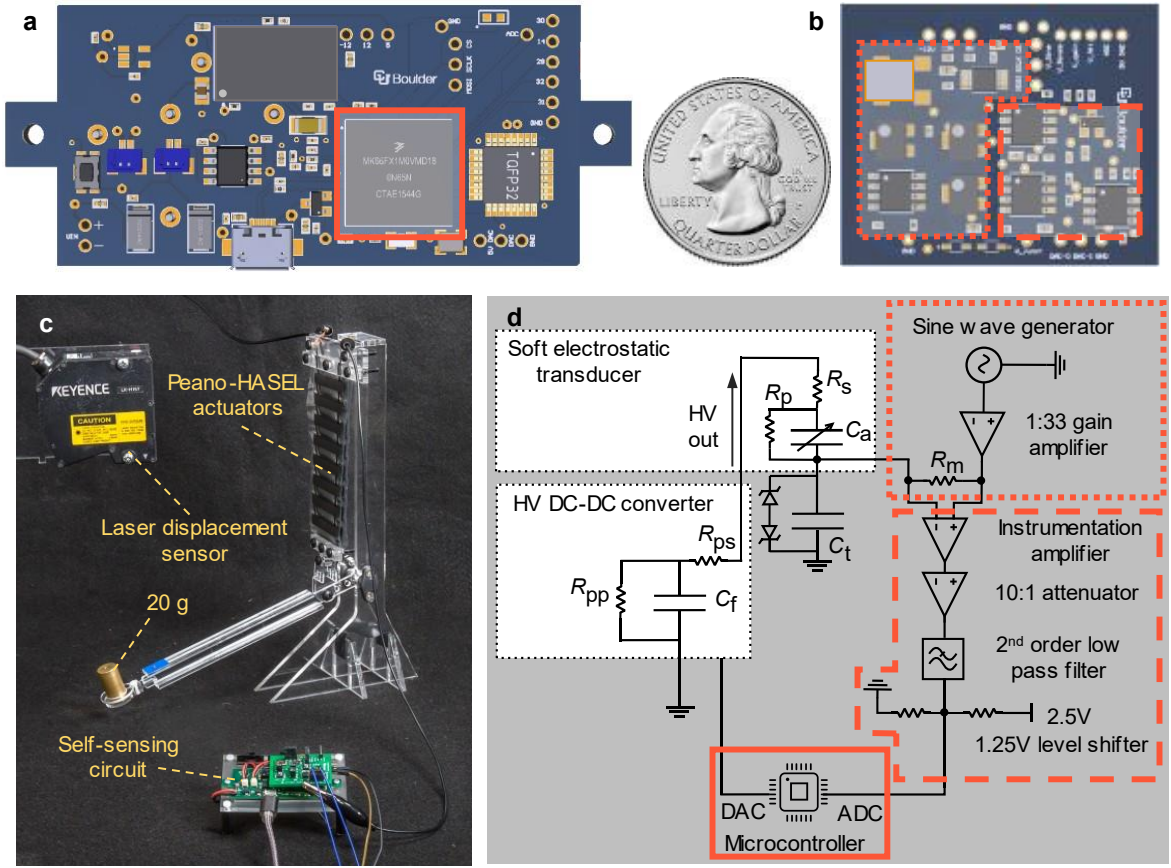

**FIG. 1.** Design of the HV capacitive self-sensing circuit. 3D models of (a) the power and processing unit and (b) the self-sensing circuit. (c) Setup for the PID-controlled Peano-HASEL arm using self-sensing as feedback. (d) Schematic diagram of the self-sensing and control system.

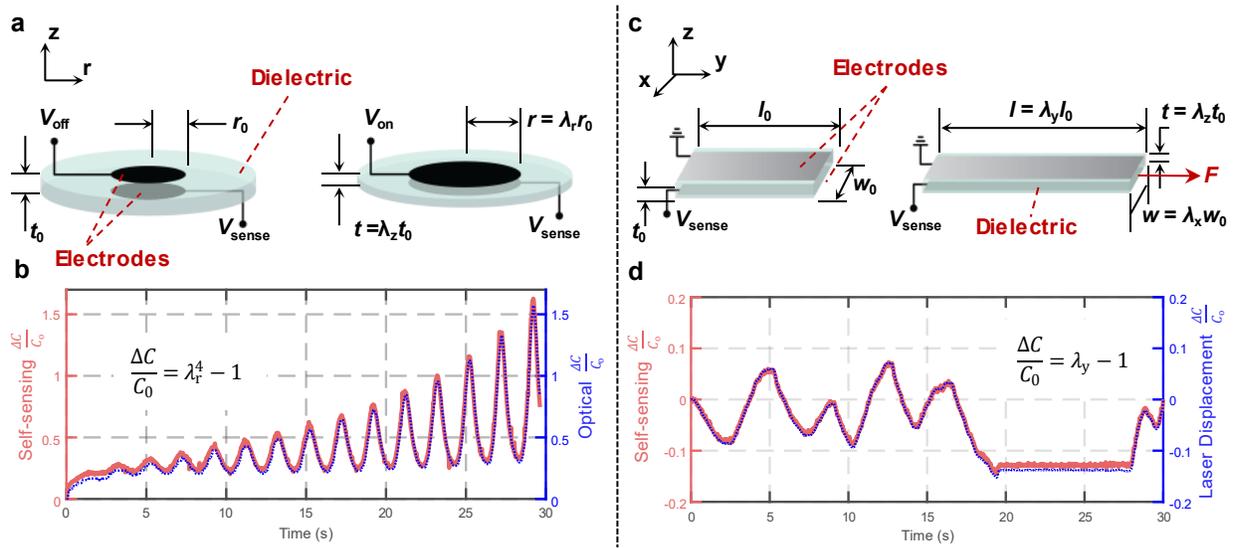

**FIG. 2.** Self-sensing of a dielectric elastomer actuator (DEA) and a stretchable capacitive sensor. (a) Schematic for a circular DEA consisting of a dielectric layer sandwiched between two stretchable electrodes. The electrodes have an initial radius $r_0$ with a dielectric thickness $t_0$ in between the electrodes. Applying voltage to the electrodes causes an increase in electrodes' radius and decrease in the thickness of the intervening dielectric layer. (b) Comparing the relative change in capacitance, $\Delta C/C_0$, using our self-sensing circuit, to relative change in capacitance using optically measured radial stretch, $\lambda_r$, showed good agreement between the two measurement methods. (c) Schematic for a stretchable sensor with initial length $l_0$, width $w_0$, and thickness $t_0$. When stretched uniaxially by force, $F$, applied in the y-direction, the sensor increases in length while decreasing in thickness and width. (d) The relative change in capacitance, $\Delta C/C_0$, using our self-sensing circuit is compared with the relative change in capacitance using measured linear stretch, $\lambda_y$, which is obtained by a laser displacement sensor.

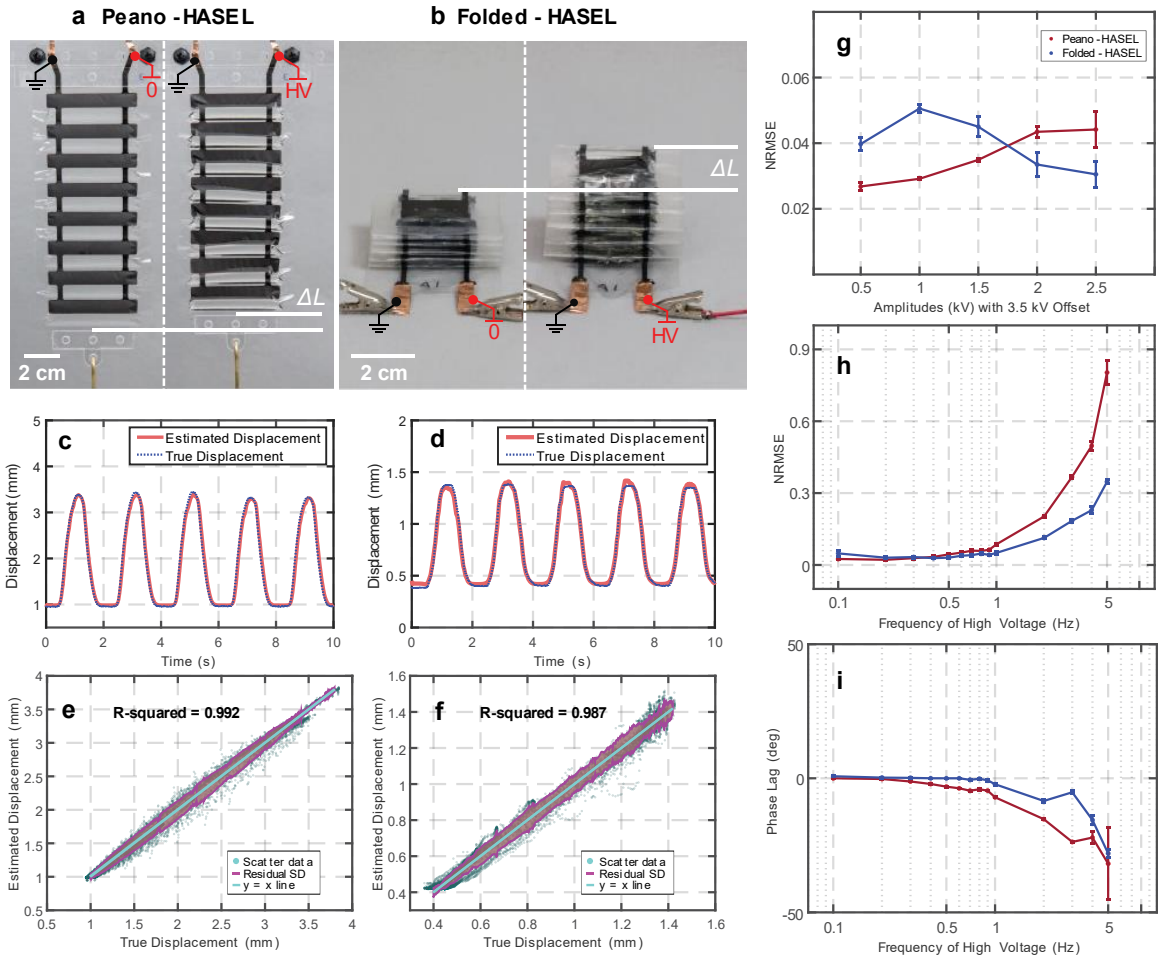

**FIG. 3.** Offline characterization, in which the self-sensing data are collected then later analyzed, for two types of HASEL actuators. (a) Peano-HASEL actuator. (b) Folded-HASEL actuator. Typical displacement results based on 5 cycles of high-voltage (HV) actuation at 0.5 Hz, with 2.5 kV amplitude and 3.5 kV offset for (c) a Peano-HASEL actuator and (d) a folded-HASEL actuator. (e) Scatter plot of estimated vs true displacements for a Peano-HASEL actuator using calibration data from all HV amplitudes and frequencies up to 1 Hz. (f) Scatter plot of estimated vs true displacements for a folded-HASEL actuator using calibration data from all HV amplitudes and frequencies up to 1 Hz. The plots include residual standard deviation bounds. (g) RMSE normalized by the actuators' displacement range as a function of HV amplitudes at 0.5 Hz. (h) RMSE normalized by the actuators' displacement range as a function of HV signal frequencies at 2.5 kV amplitude and 3.5 kV offset. (i) Phase lag between the estimated displacement and the true displacement as a function of different HV signal frequencies. Error bars show one standard deviation from the mean across ten trials.

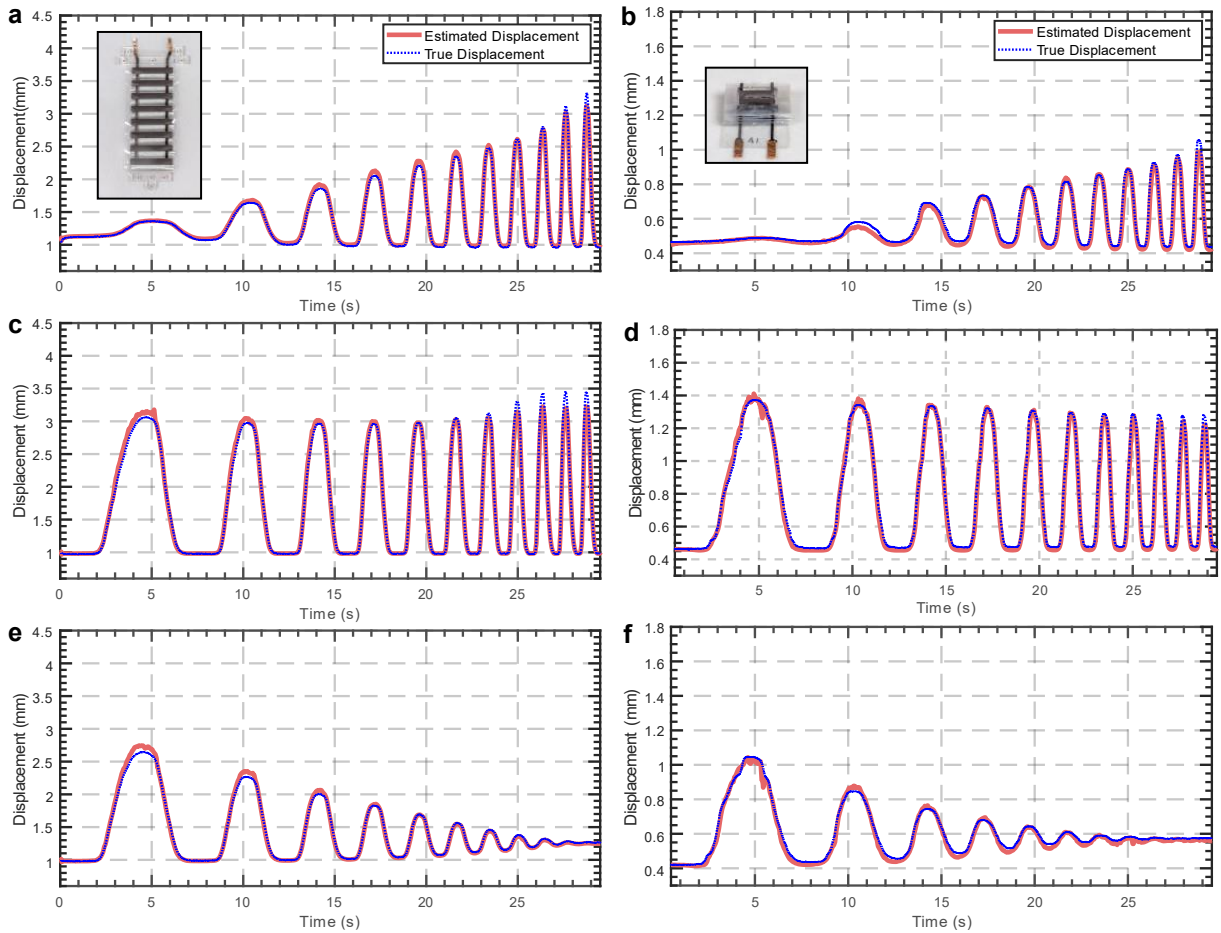

**FIG. 4.** Real-time validation of self-sensing for a Peano-HASEL actuator and a folded-HASEL actuator using a Trek 50/12 voltage amplifier. Estimated displacements from self-sensing with respect to 3.25 kV offset sinusoidal HV with both increasing amplitudes from 0.5 kV to 2.25 kV and frequencies from 0.1 Hz to 2 Hz for (a) a Peano-HASEL actuator and (b) a folded-HASEL actuator, 2.25 kV amplitude and increasing frequencies from 0.1 Hz to 2 Hz for (c) a Peano-HASEL actuator and (d) a folded-HASEL actuator, and decreasing amplitudes from 2.25 kV to 0.5 kV and increasing frequencies from 0.1 Hz to 2 Hz for (e) a Peano-HASEL actuator and (f) a folded-HASEL actuator. The HV signals' offset and amplitude are chosen to account for the actuators' activation voltage at 1 kV and failure voltages of 6 kV for Peano HASEL actuators and 10 kV for Peanut HASEL actuators.

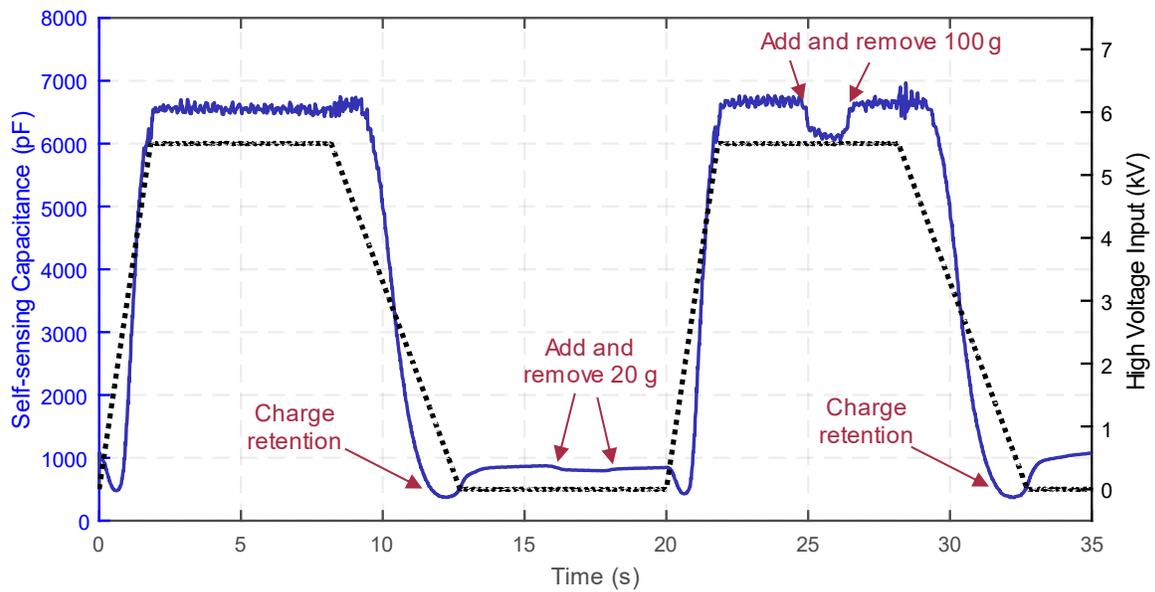

**FIG. 5.** Demonstration of self-sensing sensitivity with a Peano-HASEL actuator. An asymmetric HV ramp input at 0.05 Hz from 0 kV to 5.5 kV is applied to the actuator for 5 cycles, two of which are shown. The self-sensing circuit detects a 20 g weight applied to the actuator when no HV is applied and 100 g of weight on the actuator when 5.5 kV is applied. The effects of charge retention in HASEL actuators can be clearly detected by the self-sensing circuit – this leads to full actuator relaxation (lowest capacitance) at low voltages of ~ 1 kV, and a slight reactivation (e.g. contraction and slightly higher capacitance) at 0 kV. This effect is observed in measured displacement (see Video S2).

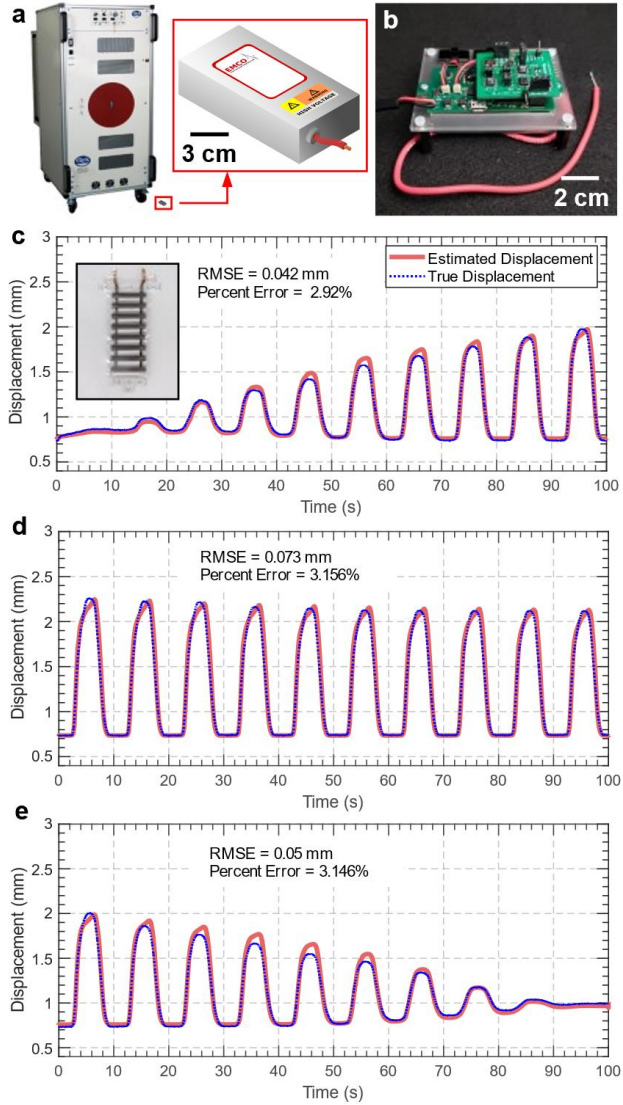

**FIG. 6.** Capacitive self-sensing of a Peano-HASEL actuator with a miniature EMCO C80 DC-DC converter. (a) 3D render of the EMCO C80 with a Trek 50/12 shown for comparison of size. (b) Fully-integrated and miniaturized power and self-sensing circuit with EMCO C80 connected underneath (out of view). Comparison between the estimated displacements of the actuator using the self-sensing circuit and the true displacement of the actuator using a laser displacement sensor with respect to high voltage with (c) increasing amplitude from 0 kV to 1.5 kV, (d) constant amplitude at 1.5 kV, and (e) decreasing amplitude from 1.5 kV to 0 kV. A frequency of 0.1 Hz and DC offset of 2.3 kV were used.

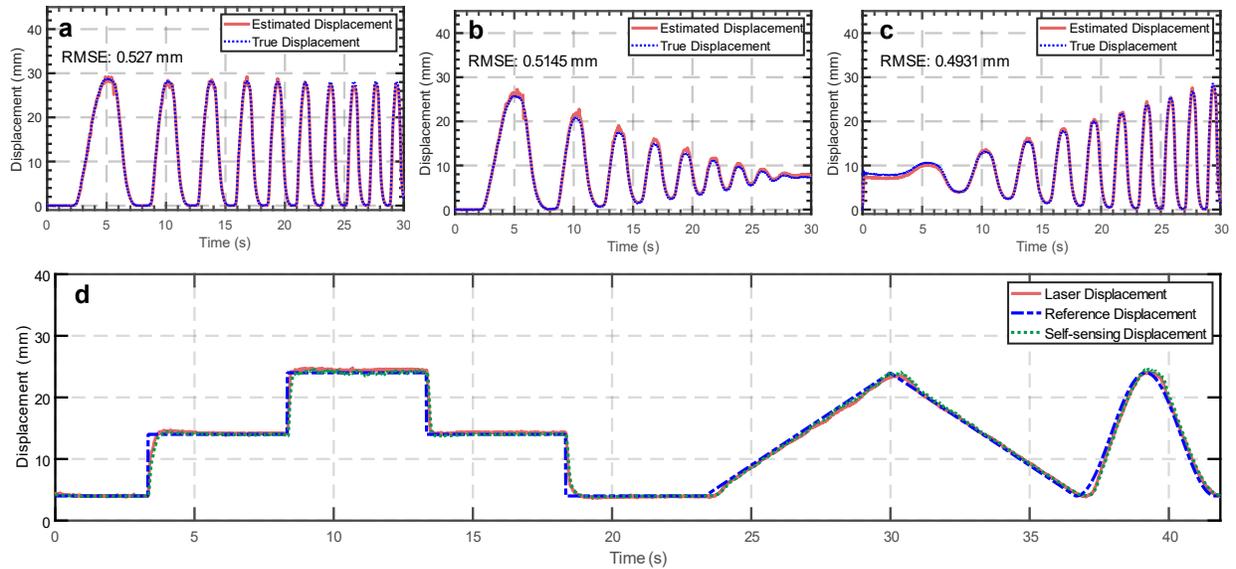

**FIG. 7.** Offline characterization, in which the self-sensing data is collected then later analyzed, for a robotic arm driven by two Peano-HASEL actuators (Figure 1c) lifting a 20-g load. The characterization process is the same as Figure 3. Estimated displacements from self-sensing with respect to a 3.25 kV offset sinusoidal HV with (a) constant 2.25 kV amplitude and increasing frequencies from 0.06 Hz to 0.6 Hz, (b) decreasing amplitudes from 2.25 kV to 0 kV and increasing frequencies from 0.06 Hz to 0.6 Hz, and (c) both increasing amplitudes from 0 kV to 2.25 kV and frequencies from 0.06 Hz to 0.6 Hz. (d) Demonstration of closed-loop control of the robotic arm using a PID controller self-sensing feedback. There are three reference displacements for control: a sequence of steps from 4 mm to 16 mm, 24 mm, 16 mm, then to 4 mm, with a 5-second hold at each step, a 3 mm/s ramp from 4 mm to 24 mm, and one cycle of a 0.2 Hz sinusoidal wave with 14 mm offset and 10 mm amplitude. Reference displacements are the desired displacements generated from the microcontroller.

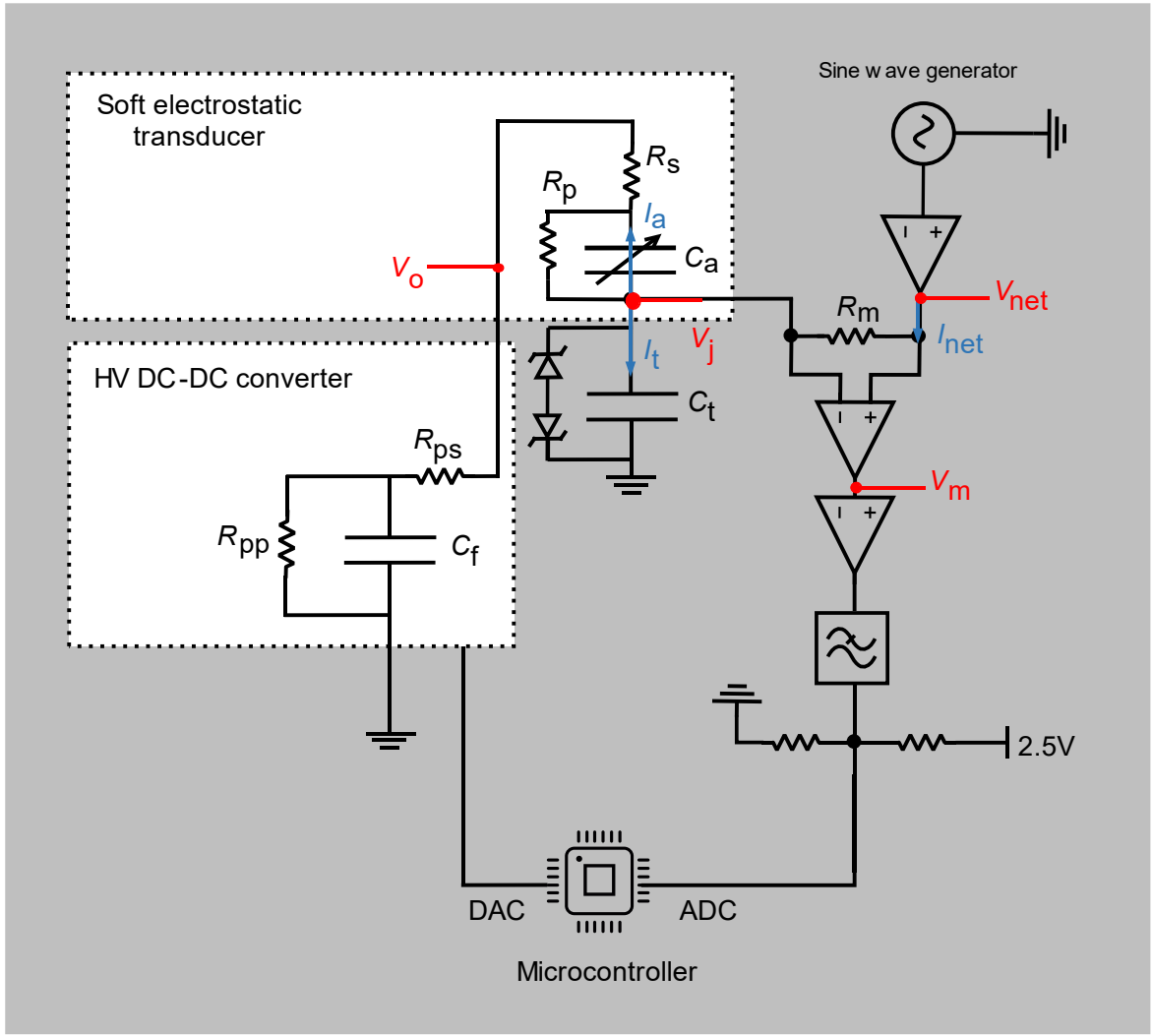

**FIG. S1.** Schematics of the self-sensing circuit with AC signal labeling. $V_{net}$ is the total voltage output from the sine wave generator; this value is chosen by adjusting the gain of the operational amplifier. $V_m$ is the voltage across the measuring resistor; this value is measured using an instrumentation amplifier. $V_j$ is the junction voltage, which is also the voltage across the transient suppression capacitor. $V_o$ is the voltage due to the sine wave generator at the output of the HV amplifier.

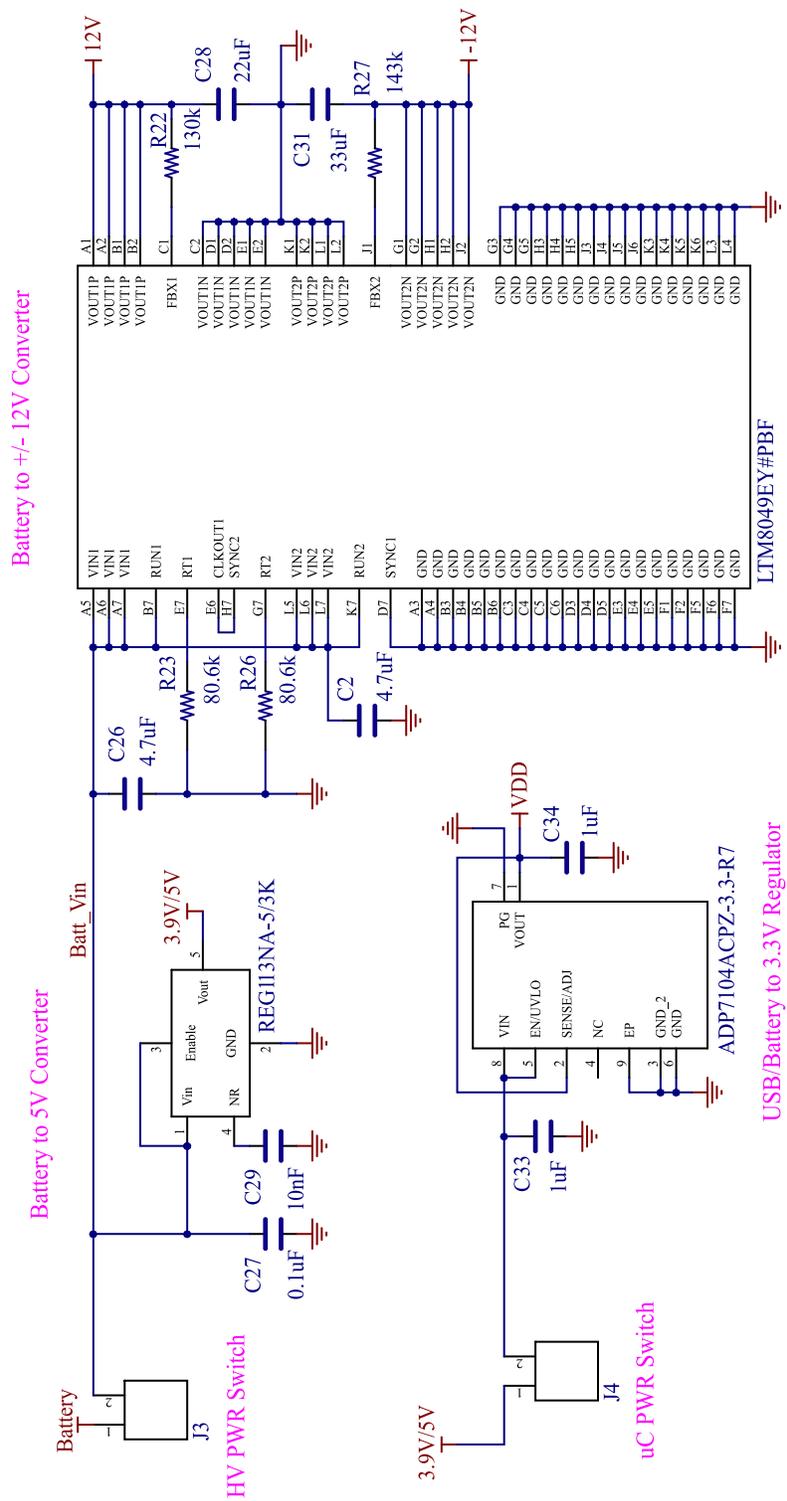

**FIG. S2.** The schematic of the power unit.

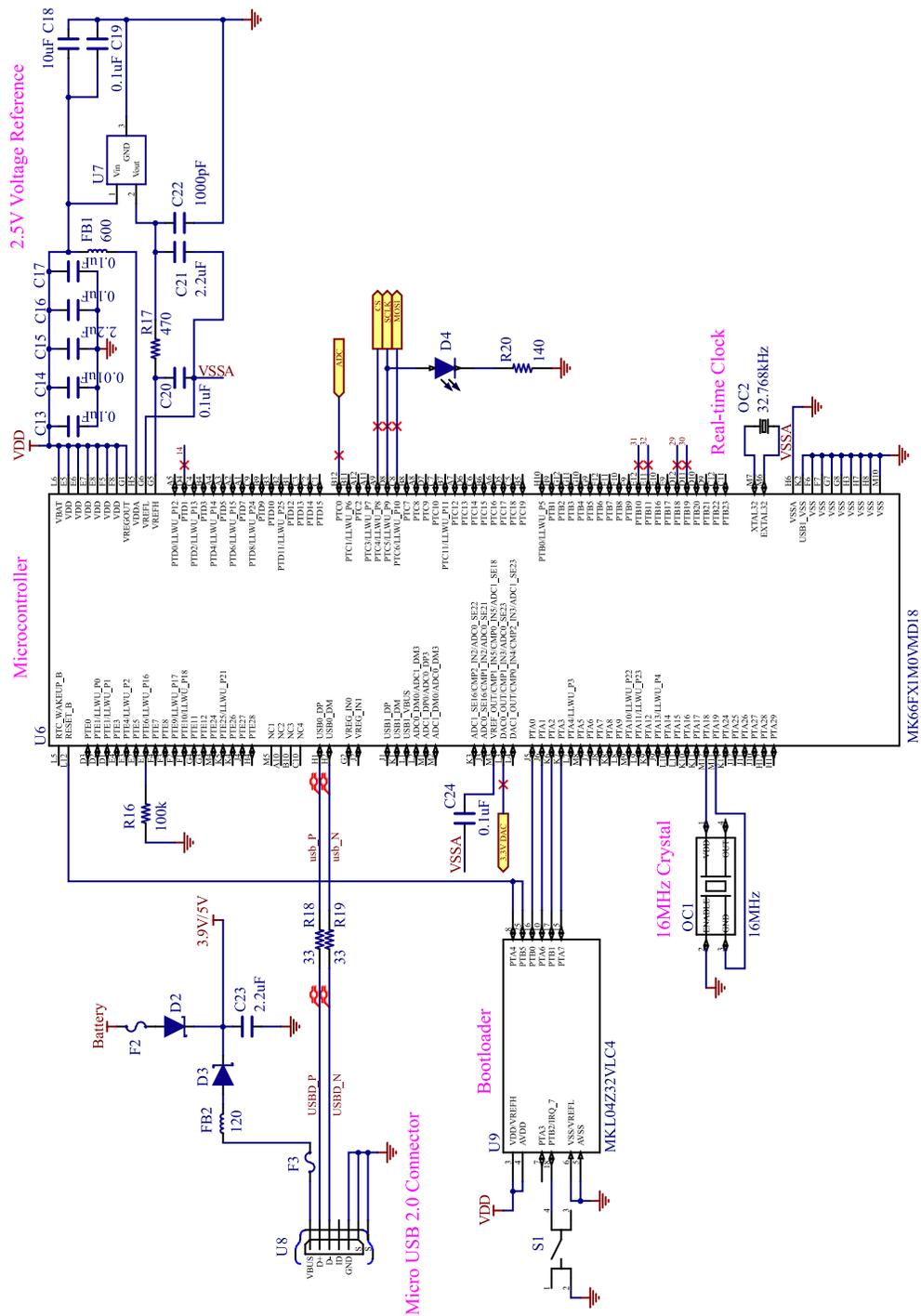

**FIG. S3.** The schematic of the processing unit. The design is based on Teensy 3.6 microcontroller from PJRC. The power and processing units are on the same PCB.

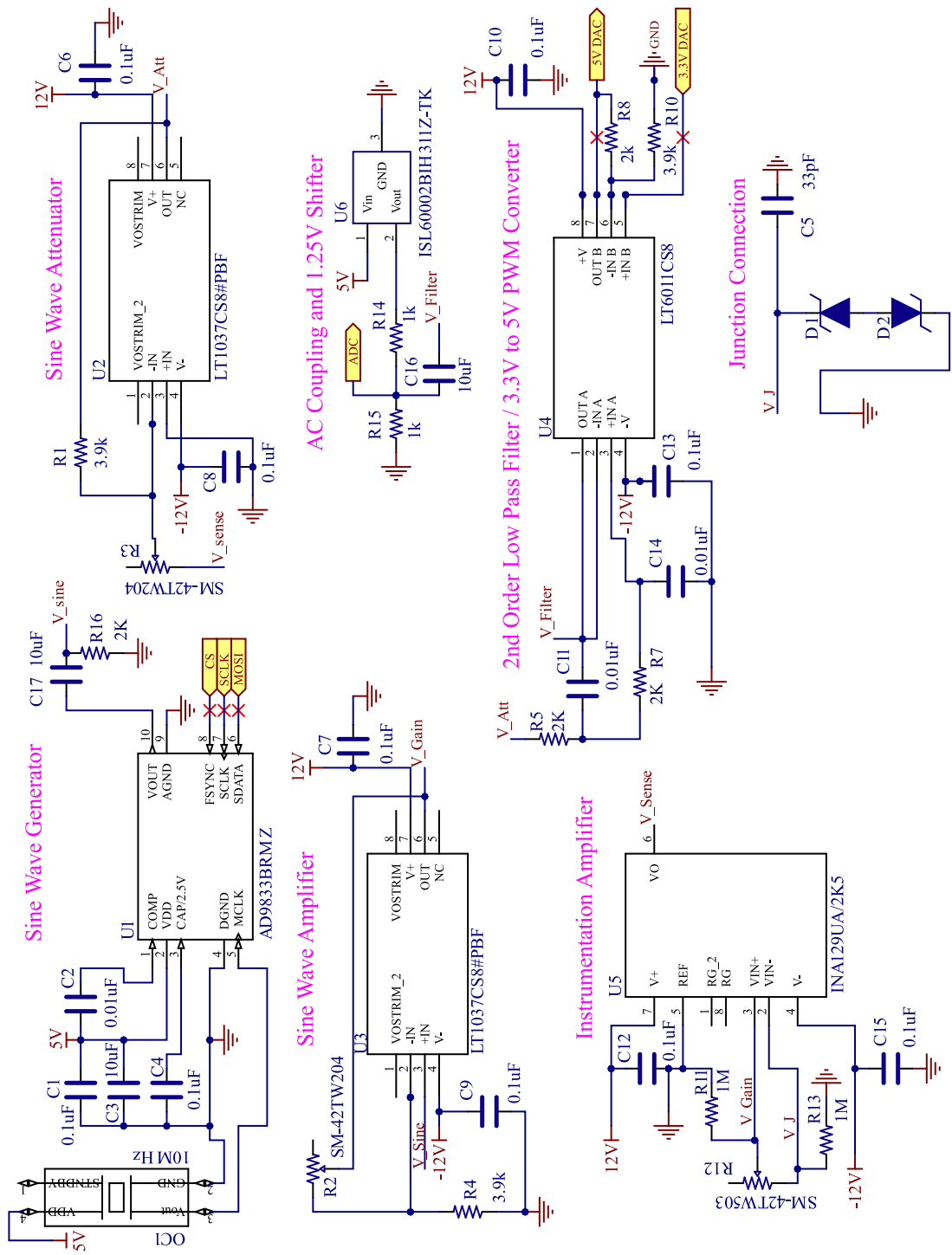

**FIG. S4.** The schematic of the self-sensing unit.

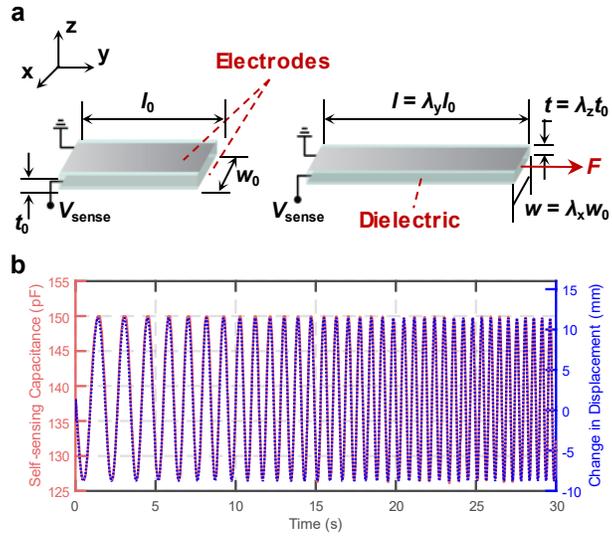

**FIG. S5.** Qualitative evaluation of phase lag between self-sensing capacitance and measured displacement of a stretchable capacitive sensor. (a) Geometric model of the capacitive sensor used for measurements. Using a dual-mode muscle lever (Model 310C, Aurora Scientific), a force, *F*, generates a sinusoidal displacement of amplitude 10.125 (mm) and offset 1.375 (mm) with an increasing frequency (0.5 Hz to 3 Hz) to the sensor. (b) The resulting change in capacitance was monitored using our self-sensing circuit and compared with linear displacement measurements performed simultaneously by the muscle lever. There is no noticeable delay between the self-sensing capacitance and the relative change in theoretical capacitance.

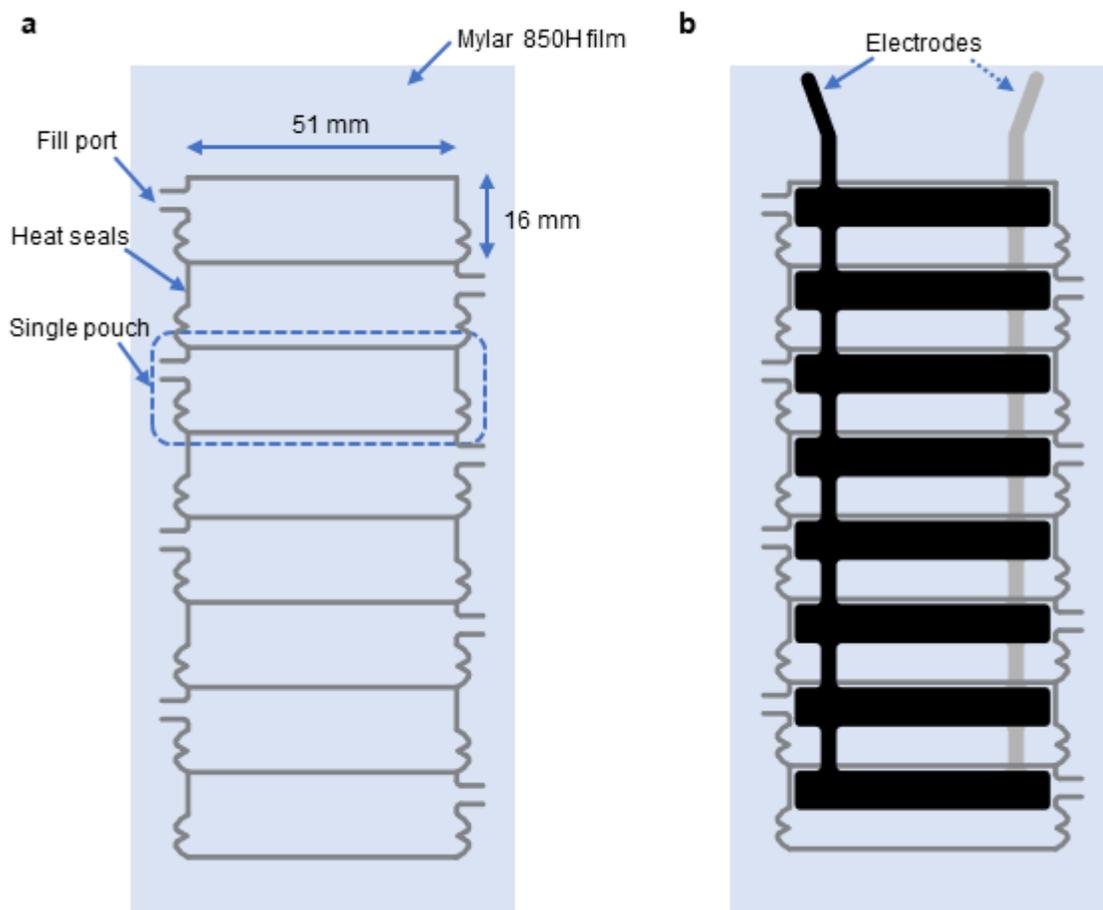

**FIG. S6.** Simplified representation of the Peano-HASEL actuator. (a) Schematic of the heat-seal pattern used in the Peano-HASEL actuators showing all eight discrete pouches and their dimensions. (b) Electrodes are screen-printed such that they cover the top half of each pouch on both sides of the actuator. Small leads at the top allow for external electrical connections to the actuator.

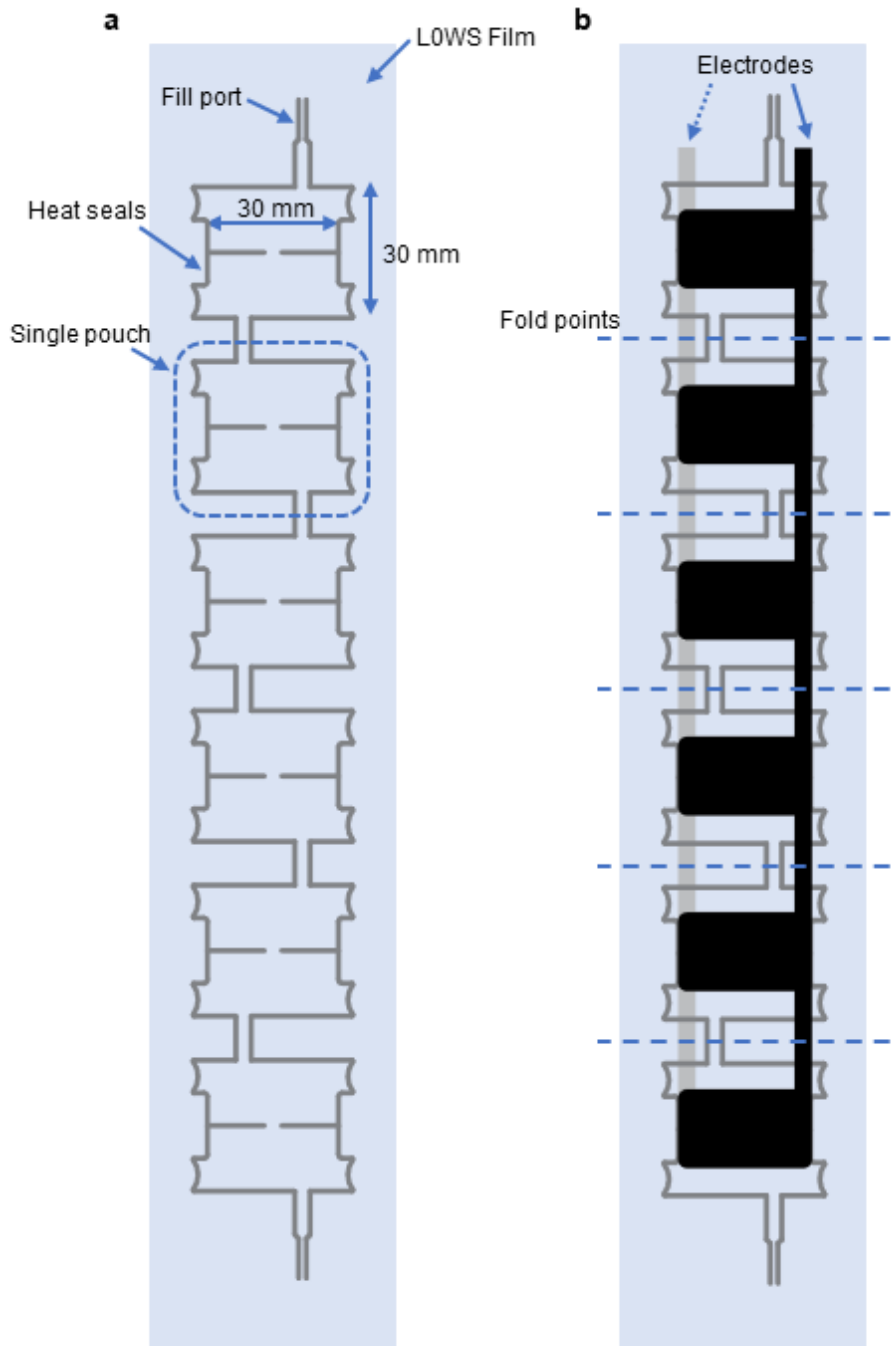

**FIG. S7.** Simplified representation of the folded-HASEL actuator (only six pouches are shown in the figure for simplicity). (a) Schematic of the heat-seal pattern used in the folded-HASEL actuators showing six of the 11 pouches and their dimensions. All pouches are initially connected to facilitate filling with liquid dielectric. (b) Electrodes are screen-printed such that they cover the middle of each pouch on both sides of the actuator. Small leads at the top allow for external electrical connections to the actuator.